\documentclass[fleqn,11pt]{article}
\usepackage{amsmath,amssymb}
\usepackage{graphicx}

\newcommand{\vfi}{\varphi}
\newcommand{\eps}{\varepsilon}
\newcommand{\R}{\mathbb{R}}
\newcommand{\C}{\mathbb{C}}
\newcommand{\dee}{{\mathrm d}}
\newcommand{\imag}{{\mathrm i}}
\newcommand{\rem}[1]{}
\newcommand{\sign}{{\rm sign}}

\begin{document}

\title{(Vanishing) Twist in the Saddle-Centre and Period-Doubling Bifurcation}
\author{H.~R.~Dullin$^{1,2}$, A.~V.~Ivanov$^{1}$
\thanks{This research is funded by the EPSRC under contract GR/R44911/01. 
Partial support by the European  Research Training Network
{\it Mechanics and Symmetry in Europe\/} (MASIE), HPRN-CT-2000-00113,
is also gratefully acknowledged. AVI was also supported in part by INTAS grant 00-221, 
RFBI grant 01-01-00335 and RFME grant E00-1-120.}\\
\\ $^{1}$ Department of Mathematical Sciences, 
\\ Loughborough University,  LE11 3TU, UK
\\ $^{2}$ Fachbereich 1, Physik, Universit\"at Bremen
\\ 28334 Bremen, Germany
\\ {\small H.R.Dullin@lboro.ac.uk},  {\small A.V.Ivanov@lboro.ac.uk}
}
\date{\today}
\maketitle

\begin{abstract}
The lowest order resonant bifurcations of a periodic orbit of a Hamiltonian system 
with two degrees of freedom have frequency ratio $1:1$ (saddle-centre) 
and $1:2$ (period-doubling).  The twist, which is the derivative of the 
rotation number with respect to the action, is studied near these bifurcations. 
When the twist vanishes the nondegeneracy condition 
of the (isoenergetic) KAM theorem is not satisfied, 
with interesting consequences for the dynamics.
We show that near the saddle-centre bifurcation the twist always vanishes.
At this bifurcation a ``twistless'' torus is created, when the resonance is passed. 
The twistless torus replaces the colliding periodic orbits in phase space.
We explicitly derive the position of the twistless torus depending on the resonance 
parameter, and show that the shape of this curve is universal.
For the period doubling bifurcation the situation is different.
Here we show that the twist does not vanish in a neighborhood of the bifurcation.


\vspace*{1ex}
\noindent
Keywords: Twist Maps; Hamiltonian Systems; Saddle-Centre Bifurcation;
Period-doubling Bifurcation; KAM; Normal Forms; Elliptic Integrals

\end{abstract}


\section{Introduction}

The dynamics near a periodic orbit of a Hamiltonian system can be 
studied in terms of a local Poincar\'e section transversal to the orbit. 
In two degrees of freedom the first return map restricted to the
surface of constant energy is an area preserving map with a fixed (or periodic) point.
If the multipliers $\mu_i$, $i=1,2$ of the fixed point have modulus 1 (but are not equal to $\pm 1$)
 the  fixed point is called elliptic. Then $\mu_1 = \overline{\mu_2} \in \C$ 
and we can write $\mu = \exp(2\pi i\omega)$ with the rotation number
$0<\omega<1/2$ of the periodic orbit.
If the periodic orbit is elliptic and the rotation number is irrational
the map can (formally) be transformed to Birkhoff normal form
which in action-angle variables $(\vfi,I)$ reads
\begin{equation} \label{eqn:BNF}
   (\vfi,I)\to (\vfi+2\pi\Omega(I),I)\,.
\end{equation}
The action $I$ is like the radial coordinate in polar coordinates, 
hence the map in normal form maps circles to circles by rotating 
them $\Omega(I)$ times.
The {\em rotation number} (or winding number) $\Omega(I)$ 
near the periodic orbit can be expanded as
\[
   \Omega(I) = \omega +  \tau_{0}I + \frac{1}{2}\tau_1 I^2+...
\]
The {\em twist} (or torsion) $\tau(I)$ is the derivative of the rotation number with 
respect to the action, 
\[
   \tau(I) =  \frac{{\rm d}\Omega}{{\rm d} I}(I)=\tau_{0}+\tau_{1}I+...
\]
When the rotation number is a strictly monotone function of the action
in some interval the map (\ref{eqn:BNF}) restricted to the corresponding 
annulus is called a {\em monotone twist map}. 
Moser's KAM theorem \cite{SM71} states that the invariant torus $I=I_0$ 
of (\ref{eqn:BNF}) persist under perturbation when its frequency $\Omega(I_0)$ 
is diophantine and its twist $\tau(I_0)$ does not vanish.
Arnold's KAM theorem \cite{Arnold78} is the same statement for flows where 
the nonvanishing of the twist corresponds to the isoeneregetic nondegeneracy 
conditions.
A well known corollary of the KAM theorem is the stability of an elliptic fixed point
in two degrees of freedom when $\Omega(0) = \omega \not = 1/3, 1/4$ and the twist at the 
origin is non-vanishing, $\tau(0) = \tau_0 \not = 0$.

When the twist vanishes the perturbed dynamics can be more complicated. 
The stability of an elliptic point can be lost when its twist vanishes, 
see \cite{DMS98b} for an example of an unstable elliptic point with $\omega = 1/5$.
The effects of vanishing twist away from the origin was first described by 
Howard \cite{HowHoh84}, and the resulting effects have been observed in many 
examples \cite{WVCP88,Sadovskii96}. 
The probably most spectacular effect is the appearance of so-called meandering 
curves \cite{HowHoh84,HowHum95,Simo98}.
The properties of non-twist maps also show interesting behaviour under 
renormalisation \cite{CGM96} and recently it has been shown \cite{DL98} that an 
extension of the KAM theorem can also be proved in this context.
In \cite{DMS98b,Moeckel90} it was finally shown that the vanishing of twist at the
fixed point generically occurs in a one parameter family when the rotation number
of a fixed point
passes through the interval $[1/4, 1/3]$. When the twist vanishes at the fixed
point a twistless torus is created in a {\em twistless bifurcation} \cite{DMS98b}.
After creation the twistless torus passes through resonances and in this way
non-twist maps generically appear in one parameter families of area preserving maps.
The truncated resonant Birkhoff normal form shows that this twistless torus
eventually collides with a saddle-centre bifurcation that gives rise to the
period 3 orbits that collide with the fixed point when $\omega = 1/3$ \cite{DMS98b}.
Such a connection between resonance and vanishing twist can also be found 
in 4 dimensional symplectic maps \cite{DM02}.

The techniques of \cite{DMS98b} can also be applied to the higher order resonances, 
in particular for $\omega = 1/4$.
In this paper we study the two remaining generic bifurcations
at even stronger resonance $\omega = 0, 1/2$. By definition the corresponding
fixed point is not elliptic. The two cases will be denoted as the $1:1$ and 
$1:2$ resonance, or as the saddle-centre and period doubling bifurcation, respectively.
The main result is that near the saddle-centre bifurcation the twist always vanishes, 
while it does not vanish near the period doubling bifurcation. 

The method is based on the analysis of the resonant normal form, in which the 
Poincar\'e map is approximated by the time $1$ map of a one 
degree of freedom system, see e.g.~\cite{MH92}. 
This normal form is an approximation, that is local near the bifurcation
in parameter space and in phase space. At first we will completely ignore this, and 
just analyse the normal forms in the following two sections. 
In section~\ref{sec:Universality} we will 
address the problem of non-locality in phase space, and also comment on the 
effect of higher order perturbations on the twistless tori. 
We will talk of invariant tori even though the invariant curves $H(u,v)=h$ of
the normal form may not be compact. 
This will also be justified in Sec.~\ref{sec:Universality}.
Finally we treat a saddle-centre bifurcation in the H\'enon map as an example.

\section{Saddle-Centre Bifurcation}

The normal form of a Hamiltonian system with two degrees of freedom near the $1:1$ resonance has the form
\begin{equation} \label{Hsn}
   H(u,v)=\frac{v^2}{2}+\frac{u^3}{3}+\eps u\,.
\end{equation}
The coefficient of $u^3$ has been scaled so that it equals $1/3$. 
The variables $u$ and $v$ are canonically conjugate variables
on a local  transversal Poincar\'e section and $\eps$ is a parameter, typically
the energy of the original system. 
The Poincar\'e map is given by the time $1$ map of the flow of $H(u,v)$. 
If the period $T$ is large, the time $1$ map advances little. 
The rotation number of the full system is the period $1$ divided by 
the period of  the reduced one degree of freedom flow, $\Omega = 1/T$.
The more familiar $\Omega = 2\pi/T$ is obtained when the time $2\pi$ map
is taken instead of the time $1$ map, but the time $1$ map is more natural 
at least for the example of the H\'enon map we are going to discuss.
Since $\Omega$ is determined by $T$, we now study in detail the period $T$
of the one degree of freedom system given by $H$. 

\begin{figure}
\centerline{\includegraphics[width=5cm]{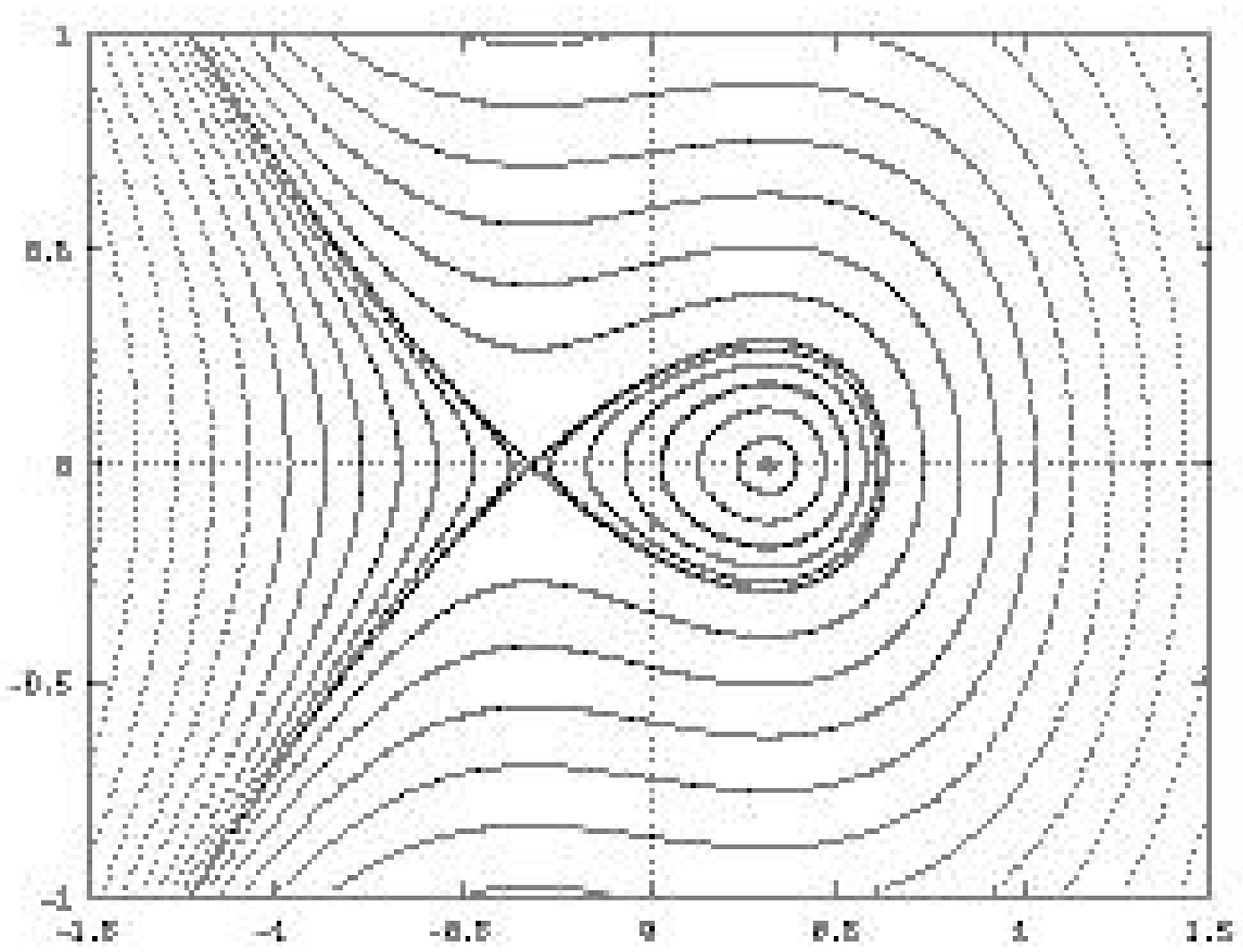}
\includegraphics[width=5cm]{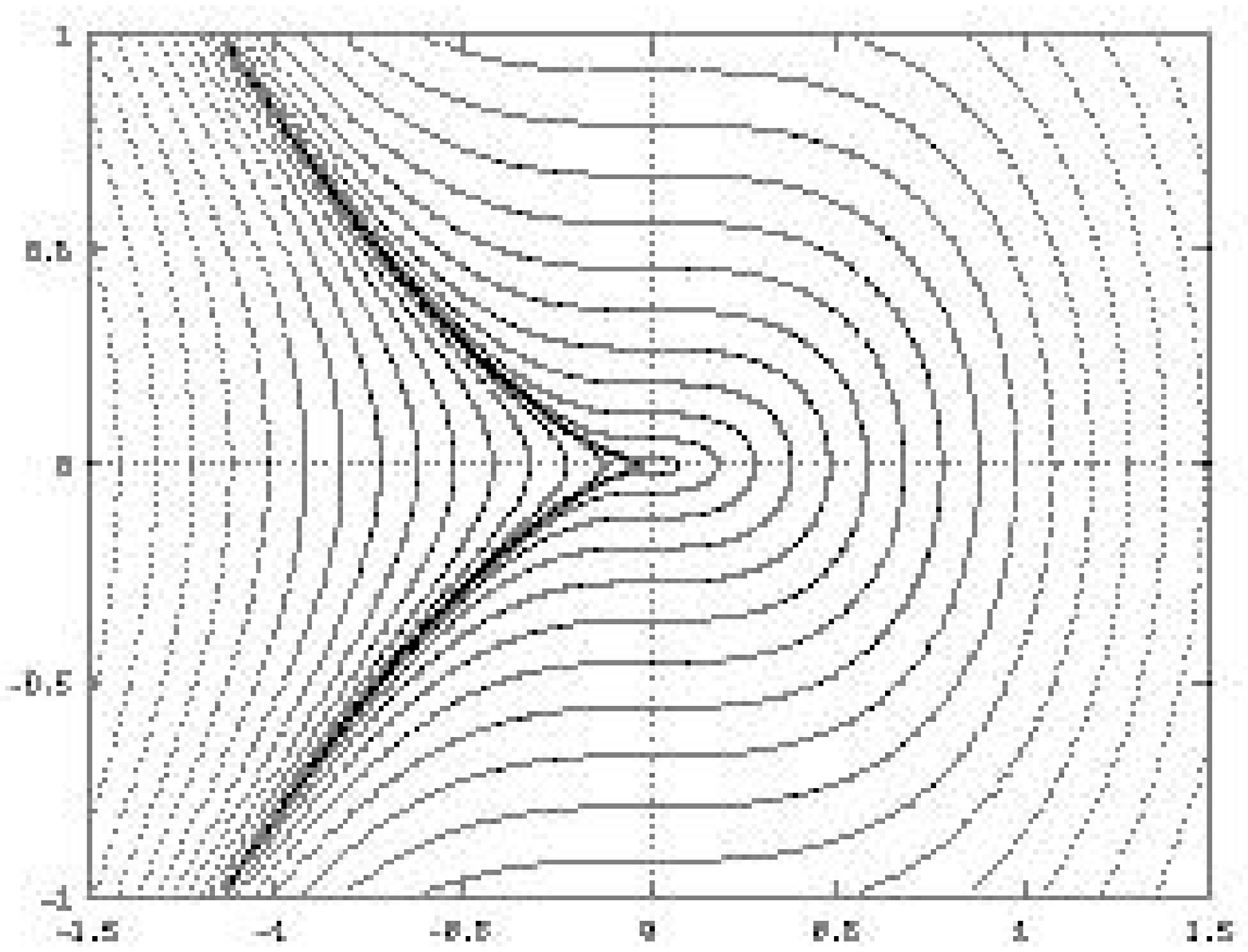}
\includegraphics[width=5cm]{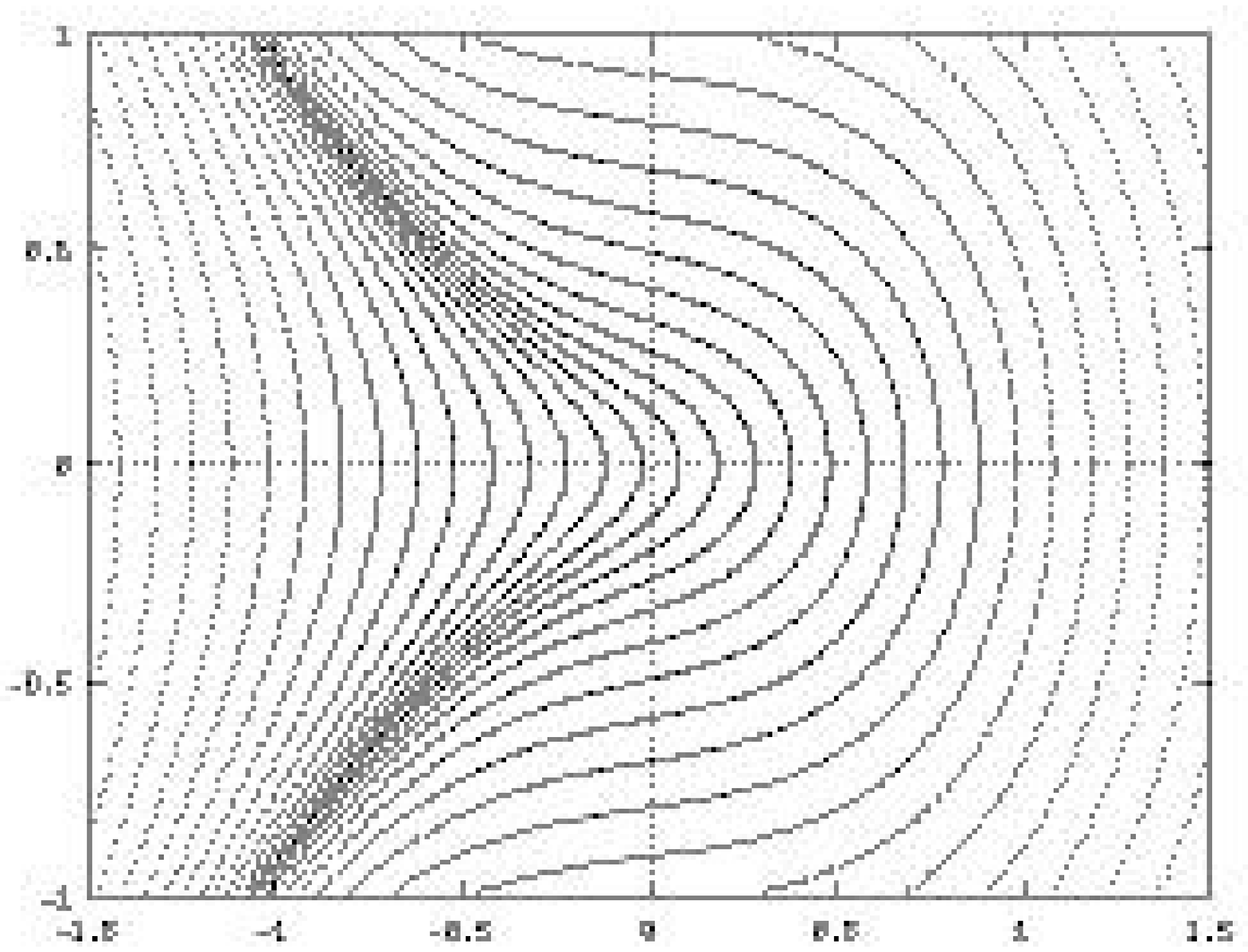}}
\caption{Lines of constant energy on the phase space $(u,v)$ for the saddle-centre 
bifurcation ($1:1$ resonance, $\omega = 0$). 
left: $\eps = -0.1$, middle: $\eps = 0$, right: $\eps = 0.1$.} \label{fig:SNphase}
\end{figure}

The critical points and critical values of the energy map $H : \R^2 \to \R$
and their dependence on $\eps$ give the main structure to the bifurcation.
Instead of a one degree of freedom system $H(u,v;\eps)$ depending on the
parameter $\eps$ one may consider $H(u,v,\eps,\theta)$ as a Hamiltonian in 
$\R^3\times S^1$, with action $\eps$ and conjugate angle $\theta$. The set of 
critical values of the energy-momentum map $(H,\eps):\R^3 \times S^1 \to \R^2$ 
is called the bifurcation diagram. A simple way to compute it is to find
the critical values of the energy map of $H(u,v)$ and consider their
parameter dependence on $\eps$.
The Hamiltonian has critical points $(u,v) = (\pm\sqrt{-\eps}, 0)$
and corresponding critical values $h=\mp2(-\eps)^{3/2}/3$. They exist when 
$\eps < 0$ and the upper sign corresponds to a local minimum of $H$, 
while the lower sign gives a saddle.
The corresponding phase portraits are shown in Fig.~\ref{fig:SNphase}.

\begin{figure}
\centerline{\includegraphics[width=5cm]{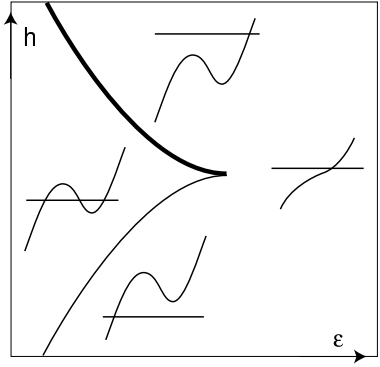}}
\caption{Schematic sketch of the bifurcation diagram for the saddle-centre bifurcation.
Graphs of $P_3$ are shown  together with a horizontal line indicating the value of $h$.
The bold lines are the critical values of the saddle. } \label{fig:SNbif}
\end{figure}

\begin{figure}
  \centerline{\includegraphics[width=9cm]{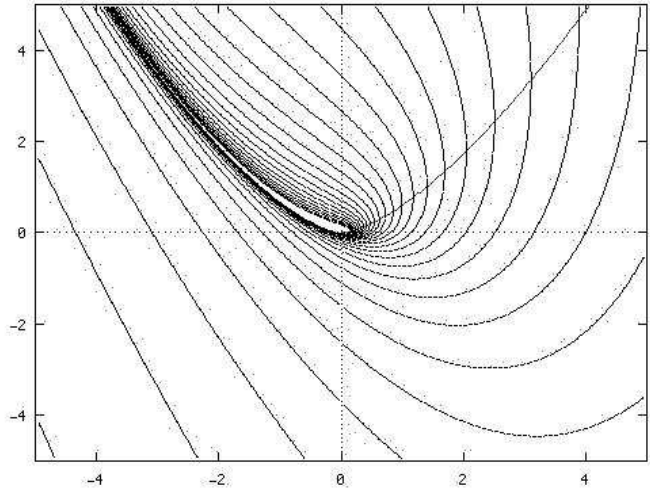}}
\caption{Lines of constant period $T=1/\Omega$ equidistant with $\Delta T = 0.3$ on the
parameter plane $(\eps, h)$ for the saddle-centre bifurcation ($1:1$ resonance, $\omega = 0$).
The vanishing twist is indicated by a curve of vertical tangents $\partial \Omega / \partial h = 0$.
}
\label{fig:SNWconst}
\end{figure}

The dynamics is given by Hamiltons equation
$\dot u = v$ and eliminating $v$ using the Hamiltonian gives a first
order equation for $u$. After separation of variables the period
of motion with energy $h$ is given by the elliptic integral
\begin{equation} \label{Teq3}
     T(h,\eps)= \oint\frac{{\rm d}u}{\sqrt{2h-\frac23 u^3-2\eps u}}\,,
\end{equation}
where the integration path is encircling the interval on the real 
axis where the argument of the square root is positive. If there are
two positive intervals either one can be taken, the result is the same.
By scaling $u=z (\sigma \eps)^{1/2}$, where $\sigma = {\rm sign}(\eps)$ 
and introducing the one essential parameter 
\begin{equation} \label{gammadef}
  \gamma = \frac{3 h}{2(\sigma\eps)^{3/2}}\,,
\end{equation}
the period $T$ is an elliptic integral on the curve
\begin{equation} \label{ElliCurve3}
  {\cal E}: w^2 = P_3(z) = 2\gamma - z^3 - 3 \sigma z\,.
\end{equation}
The case $\eps = 0$ has to be excluded in this scaling, but 
it is simple to treat it separately. We are mostly interested 
in the case where $\sigma = 1$. 
The essential integral now reads
\begin{equation} \label{Seq3}
   S(\gamma) = \oint \frac1w \, {\rm d}z
\end{equation}
and it is related to the period by
\begin{equation} \label{TSeq3}
   T(h,\eps) = \frac{\sqrt{3/2}}{ (\sigma\eps)^{1/4}}S(\gamma) \,.
\end{equation}
The polynomial $P_3$ has one or three real roots.
The collision of two real roots corresponds to the
unstable equilibrium and its separatrix. It occurs when 
the discriminant 
\[
  \Delta = -108(\sigma + \gamma^2)
\]
vanishes. $\Delta = 0$ is only possible for $\sigma = -1$
and hence the critical parameters for which a double root
occurs are given by $\gamma = \pm 1$, hence 
\begin{equation} \label{bifcurve}
  9 h^2 = -4\eps^3\,,
\end{equation}
which has a cusp at the origin. 
At the origin $h=\eps=0$ all three roots collide
in the saddle-centre bifurcation. 
The discriminant (\ref{bifcurve}) is shown in Fig.~\ref{fig:SNbif}.
In the case of the saddle-centre bifurcation the bifurcation diagram is given by 
the discriminant of $P_3$.
The bifurcation diagram divides the parameter plane $(\eps,h)$ into 
two regions: one with 3 real roots and one with 1 real and two 
complex roots. The latter has positive $\eps$ everywhere, while
the former is the wedge shaped region contained in the negative
half-plane. For $\eps < 0$ the phase portraits 
contain a pair of stable/unstable fixed points. The critical 
value of the energy of the stable fixed point is a local minimum 
given by the lower branch of the bifurcation diagram, while that 
of the unstable fixed point is a saddle given by the branch with 
positive $h$. For this range of energies there are 3 real roots.
It will turn out that the upper branch of the bifurcation diagram
corresponding to the unstable fixed point is crucial for the 
existence of vanishing twist.
For $\eps > 0$ the phase portrait is without fixed points.
Even though the topology is trivial in this case 
we will now show that the rotation number has a maximum 
on a certain invariant torus containing points near the origin 
in the phase space. At this maximum of the rotation number 
the twist vanishes. The vanishing twist occurs at the vertical
tangents of the contours of the rotation number shown in 
Fig.~\ref{fig:SNWconst}. The fact that the invariant curves 
are all unbounded for $\eps > 0$ will be dealt with in section~\ref{sec:Universality}.
For now observe that the integral (\ref{Teq3}) is finite,
even though the invariant curves are unbounded in $v$.
The main feature of the level lines of the rotation number as shown in Fig.~\ref{fig:SNWconst}
is that it 
diverges when the unstable periodic orbit is approached. This occurs
for the positive critical value of $h$ when $\eps < 0$. Everywhere else
the rotation number is a well defined, smooth and bounded function of 
$h$ and $\eps$. Accordingly the level lines ``hug'' the curve of critical 
values that correspond to the unstable orbit. Already from this property one
can deduce the existence of a curve with vanishing twist using topological arguments.
Here we proceed along the analytical route, because it will give us more
detailed information.
Note that the curve of  critical values with negative $h$, see Fig.~\ref{fig:SNbif}, 
does not appear in Fig.~\ref{fig:SNWconst}. 
One  reason  for this is that we chose to plot the rotation
number for the (non-compact) invariant tori with motion between 
$-\infty$ and the smallest real zero of $P_3$. These invariant tori do 
not contain critical points of the energy map, even though the corresponding
energy might be a critical value.
Accordingly the rotation number is smooth across this line of critical values. 
The critical point corresponding to the critical values is the stable fixed point 
at the local minimum of $H$.
But even if we would plot the rotation number of 
the bounded invariant tori near the local minimum of the potential the 
picture is unchanged. The reason is that for a cubic elliptic curve the 
integrals of first kind over either one of the two real intervals (if they exist) 
are equal.

Since $\Omega = 1/T$ and $2\pi T = \partial J/\partial h$ the derivative of
the rotation function is
\[
   \frac{\partial \Omega}{\partial J} = \frac{\partial \Omega/\partial h}{\partial J/\partial h}
      = -\frac{\partial T/\partial h}{2\pi T^3} \,.
\]
Hence the twist vanishes when
\[
  \frac{\partial T}{\partial h} =  \frac{(3/2)^{3/2}}{\eps^{7/4}} 
\frac{\partial S}{\partial \gamma}  = 0
\]
and this is only possible for finite $\eps$ when 
\[
\frac{\partial S}{\partial \gamma} = - \oint \frac{1}{w^3} \,{\rm d}z  = 0\,.
\]
This complete elliptic integral can be written as a linear combination of
Legendre's standard integrals. In this way a condition 
for the vanishing of the twist is now obtained.
The relevant case for this purpose is that of 
one real root, for which the phase portrait has no fixed point.
The integrand $w$ is positive for $u \in (-\infty, z_0)$, where
$z_0$ is the single root of $P_3(z)$. 
Let the factorized polynomial be given by
\begin{equation} \label{P3fac}
   P_3(z) = -(z-z_0)P_2(z), \qquad
   P_2(z) = (z-\zeta_1)^2 + \zeta_2^2\,,
\end{equation}
so that the complex roots are $\zeta_1 \pm i \zeta_2$.
Denote the distance between the real and complex roots by $r$, 
hence $r^2 = P_2(z_0)$, so that the discriminant is given by
$\Delta = -4r^4\zeta_2^2$.
Legendre's standard integral of the first kind $K(k)$ has differential 
\begin{equation} \label{eq:omega1}
  \omega_1 = \frac{1}{\sqrt{P_3(z)}} {\rm d}z
\end{equation}
up to a constant factor, where the modulus $k$  is given by 
\begin{equation} \label{keq}
   k^2 = \frac12 \left( 1+\frac{z_0 - \zeta_1}{r} \right) 
           = \frac12\left( 1 + \frac{\sign(\gamma)}{\sqrt{1+\alpha^2}}\right) \,,
\end{equation}
and in the last equality the parameter $\alpha = \zeta_2/(z_0-\zeta_1)$ 
has been introduced. 
In the second equality in addition $\sign( z_0 - \zeta_1) = \sign( \gamma)$ is used, 
which is true because $z_0 = \zeta_1$ in (\ref{P3fac}) together with
the vanishing of the quadratic coeffcient in (\ref{ElliCurve3}) implies $z_0 = \zeta_1 = 0$, 
and therefoe the polynomial has no constant term and $h = \gamma=0$.

A non-standard form of the differential of Legendre's standard integral
of second kind $E(k)$ is
\[
   \omega_2 =  \frac{P_2(z){\rm d}z}{(z-(z_0 \pm r))^2} \omega_1 \,,
\]
up to the same constant factor as in (\ref{eq:omega1}).
The differential $\dee z/w^3 $ we are interested in is of the second kind,
and can therefore be written as a linear combination of $\omega_1$ 
and $\omega_2$ with constant coefficients, up to a total differential:
\[
\frac{{\rm d}z}{w^3} = A \omega_1 + B \omega_2 + {\rm d} F
\]
where $F =  Q_2(z)/((z-(z_0\pm r))w)$. Together with the undetermined 
coefficients of the quadratic polynomial $Q_2$ this gives a system of 5 
linear equations for the 5 unknown coefficients. Solving these equations gives
\[
A = \frac{2r}{\Delta} ((z_0 - \zeta_1)r  - r^2 + 4\zeta_2^2), \quad
B = \frac{4r}{\Delta} (r^2 - 4\zeta_2^2)
\]

Since the quadratic coefficient of $P_3$ is zero, the roots of $P_3$
add up to zero. Therefore the real parts satisfy $z_0 + 2 \zeta_1 = 0$,
hence
\[
  2k^2 = 1 - 3\zeta_1/r \quad \mbox{and} \quad r^2 = 9 \zeta_1^2 + \zeta_2^2\,.
\]
With these equations the coefficients $A$ and $B$ can be expressed in
terms of $k$ alone, up to the factor $(r^2 \Delta)^{-1}$.
The condition of vanishing twist, $\partial S/\partial \gamma = 0$, 
finally reads
\[
(8k^4 - 9k^2 + 1)K(k)=(16k^4 - 16k^2 + 1)E(k)\, ,
\]
where $K(k)$ and $E(k)$ stand for elliptic integrals of the first and the second kind, respectively.

For $k=0$ we have equality since both elliptic integrals equal $\pi/2$. 
The first derivatives of either side vanishes, but the second derivatives
are $-35\pi/4$ and $-65\pi/4$, respectively, 
so that the left hand side is larger for small $k$.
For $k=1$ the prefactor of $K$ vanishes, while that of $E$ gives 1 
and $E(1) = 1$. Hence for $k \to 1$ the right hand side dominates.
This proves that there exists a solution of this equation
for $k \in (0,1)$. Numerically we find $k_0^2 \approx 0.7097215$.
In order to calculate the corresponding $\gamma_0$ we observe that
$\alpha$ as introduced in (\ref{keq}) is related to $\gamma$ by
\begin{equation} \label{alphaeq}
   \alpha = \frac{\zeta_2}{z_0 - \zeta_1} = -\frac{\zeta_2}{3\zeta_1}
 =  \frac{1}{\sqrt{3}}\frac{\Gamma^2 + \sigma}{\Gamma^2 - \sigma} \,,
\end{equation}
where
\begin{equation} \label{Gdef}
  \Gamma = ( \gamma + \sqrt{\sigma+\gamma^2})^{1/3} \,.
\end{equation}
Using (\ref{keq}) and $k_0$ the corresponding value of $\alpha$ is 
$\alpha_0 \approx 2.164255$, 
and from $\alpha_0$ using (\ref{alphaeq}) we find $\gamma_0 \approx 0.9152203 $.
Therefore the curve of vanishing twist in the parameter plane occurs for positive $\eps$ 
when 
\begin{equation} \label{VTg0curve}
  3 h = 2 \gamma_0 \eps^{3/2}\,.
\end{equation}
Since $\gamma_0 < 1$ the curve of vanishing twist is bent downward
as compared to the bifurcation curve (\ref{bifcurve}) for 
$\eps < 0$ and $h>0$. See Fig.~\ref{fig:SNWconst} for a graph of this
curve together with the numerically computed lines of constant rotation number.
The lines of constant rotation number have vertical slope at their intersection
with the critical curve, as must be the case.

\begin{figure}
  \centerline{\includegraphics[width=7cm]{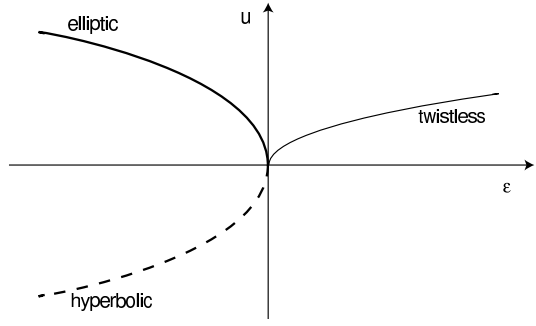}}
\caption{Bifurcation diagram of the position in phase space $u$ versus the bifurcation
parameter $\eps$. For $\eps < 0$ the location of the fixed points at $\pm \sqrt{-\eps}$ are shown, 
while for $\eps > 0$ the maximal $u$ of the twistless torus at $z_0(\gamma_0) \sqrt{\eps}$ is shown.}
\label{fig:Snubd}
\end{figure}

The scaling of $u$ reduces the number of parameters to one. 
The essential parameter $\gamma$, in its dependence on $h$ 
and $\eps$, organizes the bifurcation. It allows to compute explicitly
all the important characteristics of the twistless torus.
Combining (\ref{keq}) and (\ref{alphaeq}) shows that $k$ is a function 
of $\gamma$. The curves in the parameter plane that have the same
value of $\gamma$ are given by (\ref{gammadef}). The most prominent
ones are the curve of critical values $\gamma = \pm1$, as shown in Fig.~\ref{fig:SNbif},
and  $\gamma = \gamma_0$, the curve of twistless tori, see Fig.~\ref{fig:SNWconst}. 
They all have the same shape
of a semicubical parabola, except when $\gamma = 0$, hence $h=0$,
or $\gamma = \pm\infty$, hence $\eps = 0$ with $\pm h > 0$. 
\[
\begin{array}{cccc}
k^2                          & \gamma & \alpha & \text{curve} \\ \hline
1                               &        +1   &   0                           & 3h = -2(-\eps)^{3/2} \\
\frac14(2+\sqrt{3}) & +\infty    &   {1}/{\sqrt{3}} &  \eps = 0, h>0 \\
0.709721497  & 0.91522       &  2.164255             & 3h = \gamma_0(-\eps)^{3/2} \\
\frac12                     &   0          & \infty                       & \eps = 0, h > 0 \\
\frac14(2-\sqrt{3})  & -\infty    &   {1}/{\sqrt{3}} &  \eps = 0, h<0 \\
0                               &     -1      &  0                            & 3h = 2(-\eps)^{3/2} 
\end{array}
\]
Approaching the bifurcation point along the curve (\ref{gammadef}) gives 
$k(\gamma)$ in the limit. The function $k(h,\eps)$ is therefore 
not continuous at the origin. 
Moving on curves (\ref{gammadef}) in the parameter plane for any $\gamma \neq 1$
the change in the period $T(h,\eps)$ given by (\ref{Teq3}) is elementary. From
(\ref{TSeq3}) it follows that $T$ is proportional to $|\eps|^{-1/4}$, and the constant
of proportionality is determined from (\ref{TSeq3}) and (\ref{Seq3}).
The divergence of the period upon approaching the bifurcation is therefore
not caused by the elliptic integral, but merely by the algebraic dependence 
$|\eps|^{-1/4}$. 
In this way the rotation number of the stable periodic orbit is given by 
\[
     \Omega(\gamma=-1,\eps) =  \frac{1}{\sqrt{2}\pi} (-\eps)^{1/4}  \approx 0.225079 \, |\eps|^{1/4}\,.
\]
The integral $S(-1)$ contained in this expression can be easily calculated using 
residue calculus because for $\gamma = -1$ the curve has a double root, $P_3(z) = -(z+2)(z-1)^2$.
For general values of $\gamma$ and in particular for the twistless torus the
elliptic integral $S(\gamma)$ (\ref{Seq3}) needs to be calculated.
The single real root is given by 
\begin{equation} \label{z00}
    z_0 = \Gamma - \frac{\sigma}{\Gamma}, \quad \text{hence} \quad
   z_0(\gamma_0) \approx 0.5535942\,.
\end{equation}
Finally $S(\gamma)$ is obtained as
\[
     S(\gamma) = \frac{4K(k) }{\sqrt{3 z_0 \sqrt{1+\alpha^2}/2 } } \,.
\]
In particular when $\gamma = \gamma_0$ the rotation number 
of the twistless torus is
\begin{equation} \label{Omlim}
     \Omega(\gamma=\gamma_0, \eps) =
      \frac{ \sqrt{ z_0(\gamma_0) \sqrt{1+\alpha^2_0}}} {4K(k_0) } \eps^{1/4}
       \approx 0.1374244 \, \eps^{1/4}\,.
\end{equation}
The constant of proportionality is close to $\sqrt{3}/4\pi$. 
Using the above value of $z_0$ the position of the rightmost point
of the twistless torus in phase space is located at  $u_0 = z_0 \sqrt{\eps}$.
This means that for $\eps = \eps_0$ this point is on the same side as 
the stable periodic orbit was for $\eps  = -\eps_{0}$ before the bifurcation, 
but by a factor of $1/z_0(\gamma_0) \approx 2$ closer to the origin. 
See Fig.~\ref{fig:Snubd} for an illustration in the form 
of a standard bifurcation diagram showing position in phase space $u$
versus bifurcation parameter $\eps$.


\section{Period-Doubling Bifurcation}

\begin{figure}
\centerline{\includegraphics[width=5cm]{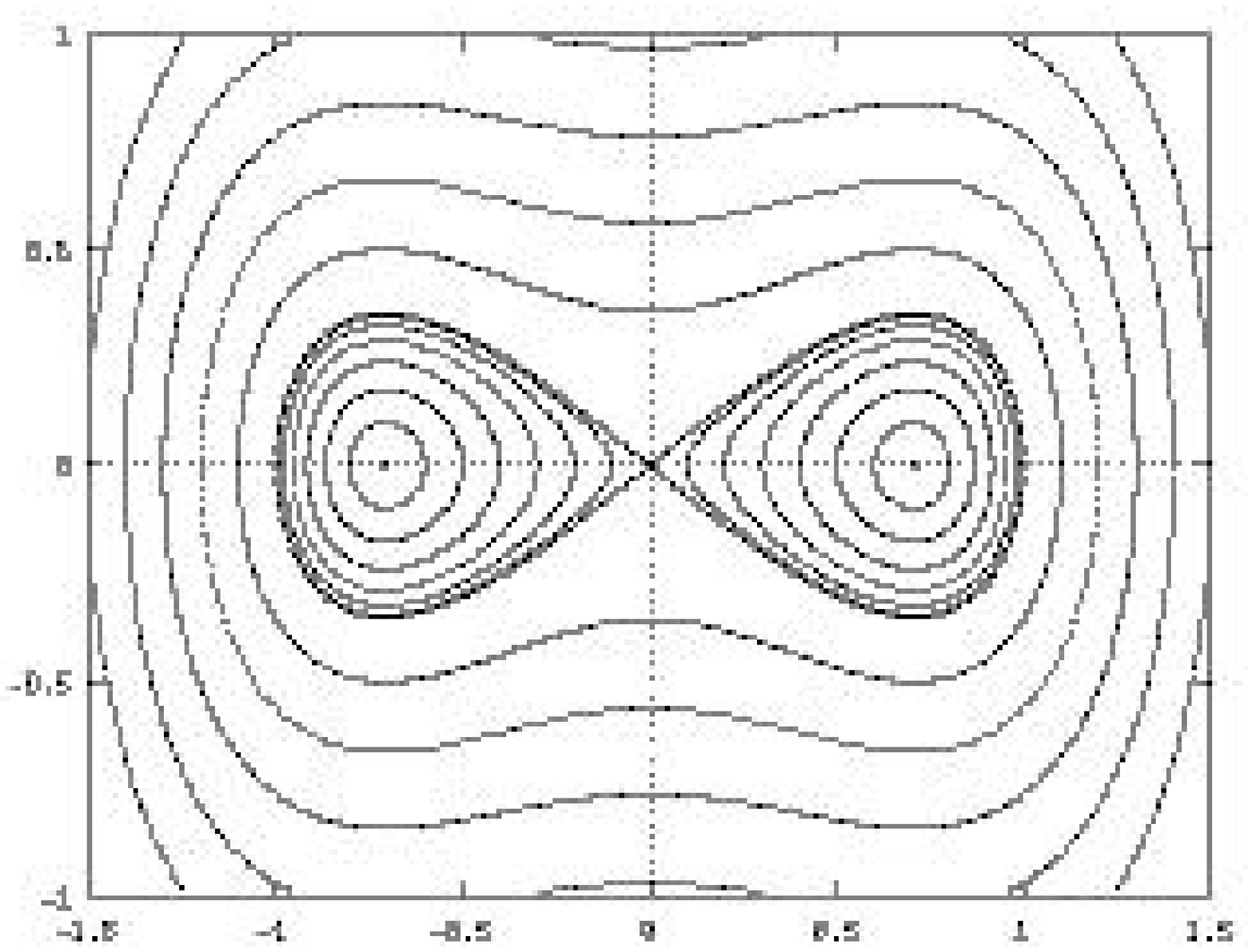}
\includegraphics[width=5cm]{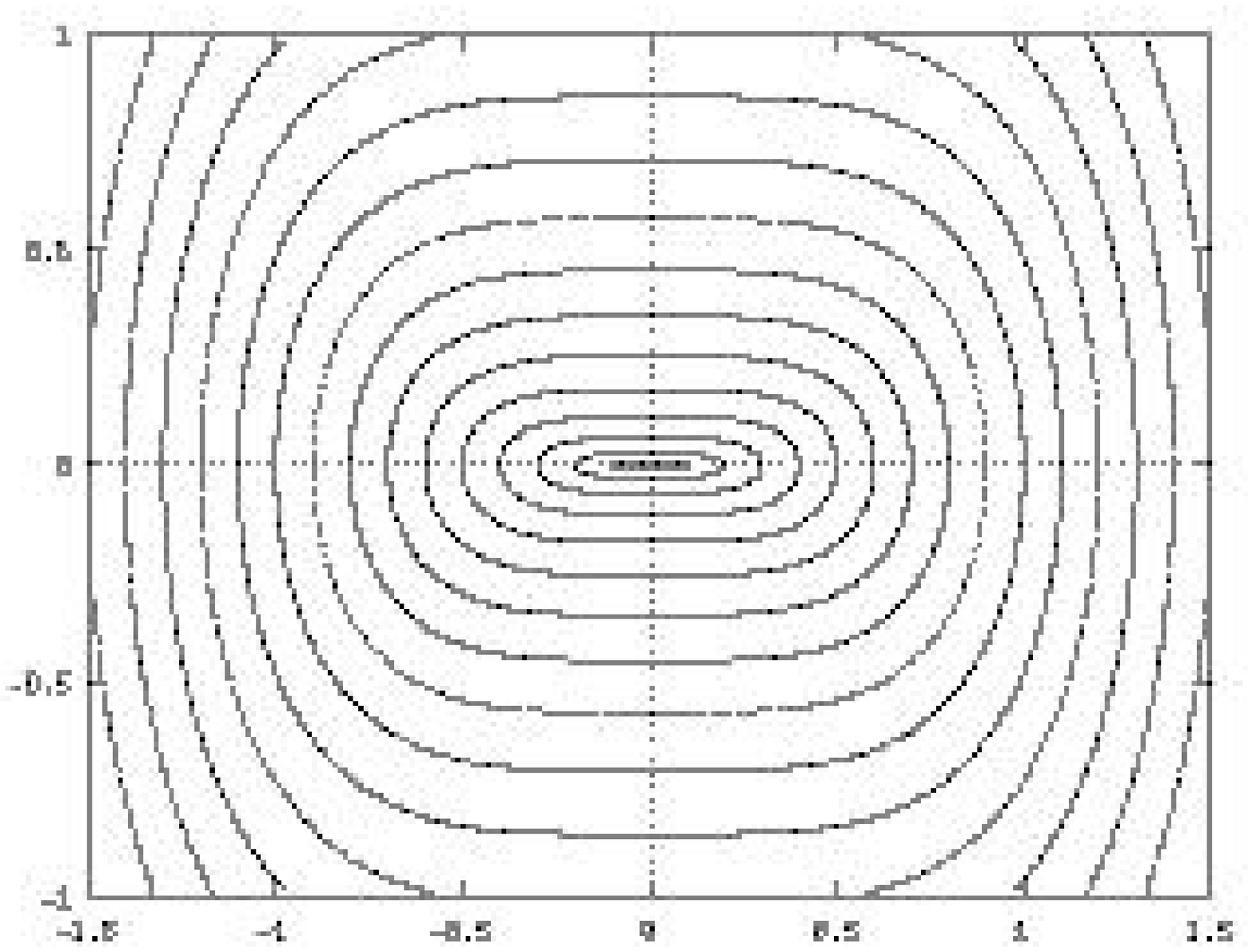}
\includegraphics[width=5cm]{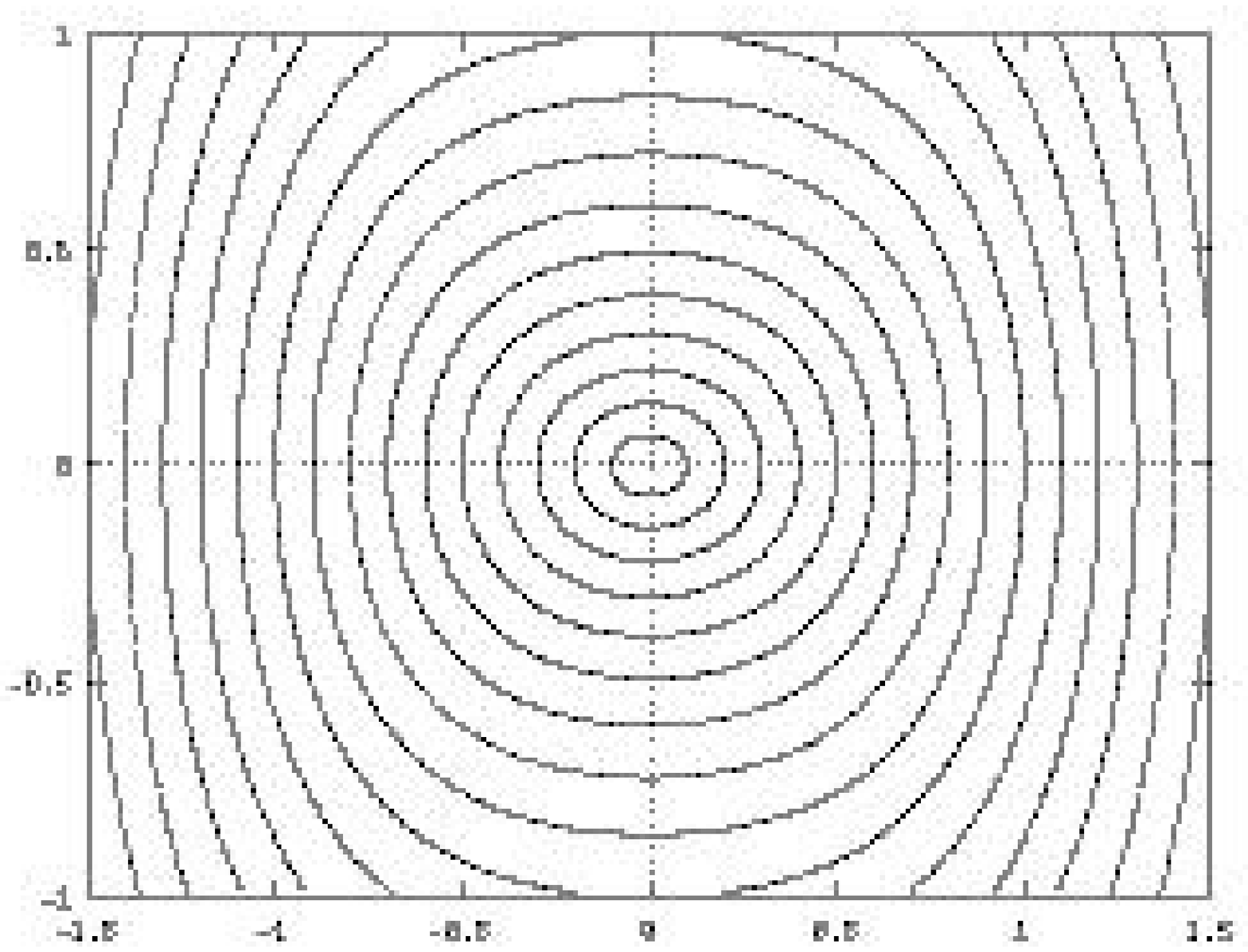}}
\caption{Lines of the constant energy on the phase space $(u,v)$ 
for the period-doubling bifurcation $(D = 1)$ for $\eps = -0.1$, $\eps = 0$, $\eps  = 0.1$.}
\label{fig:PDphaseP}
\end{figure}

\begin{figure}
\centerline{%
\includegraphics[width=5cm]{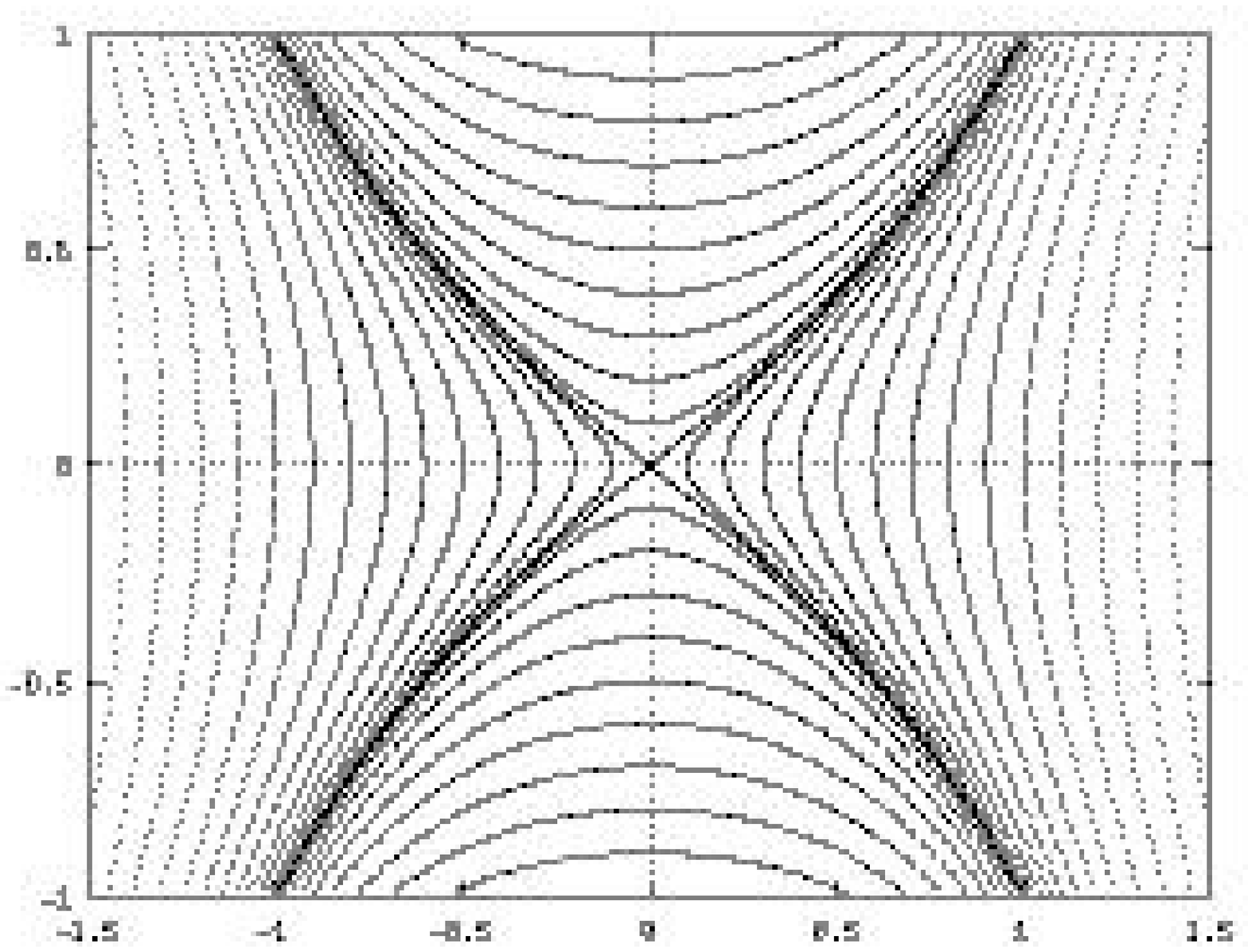}
\includegraphics[width=5cm]{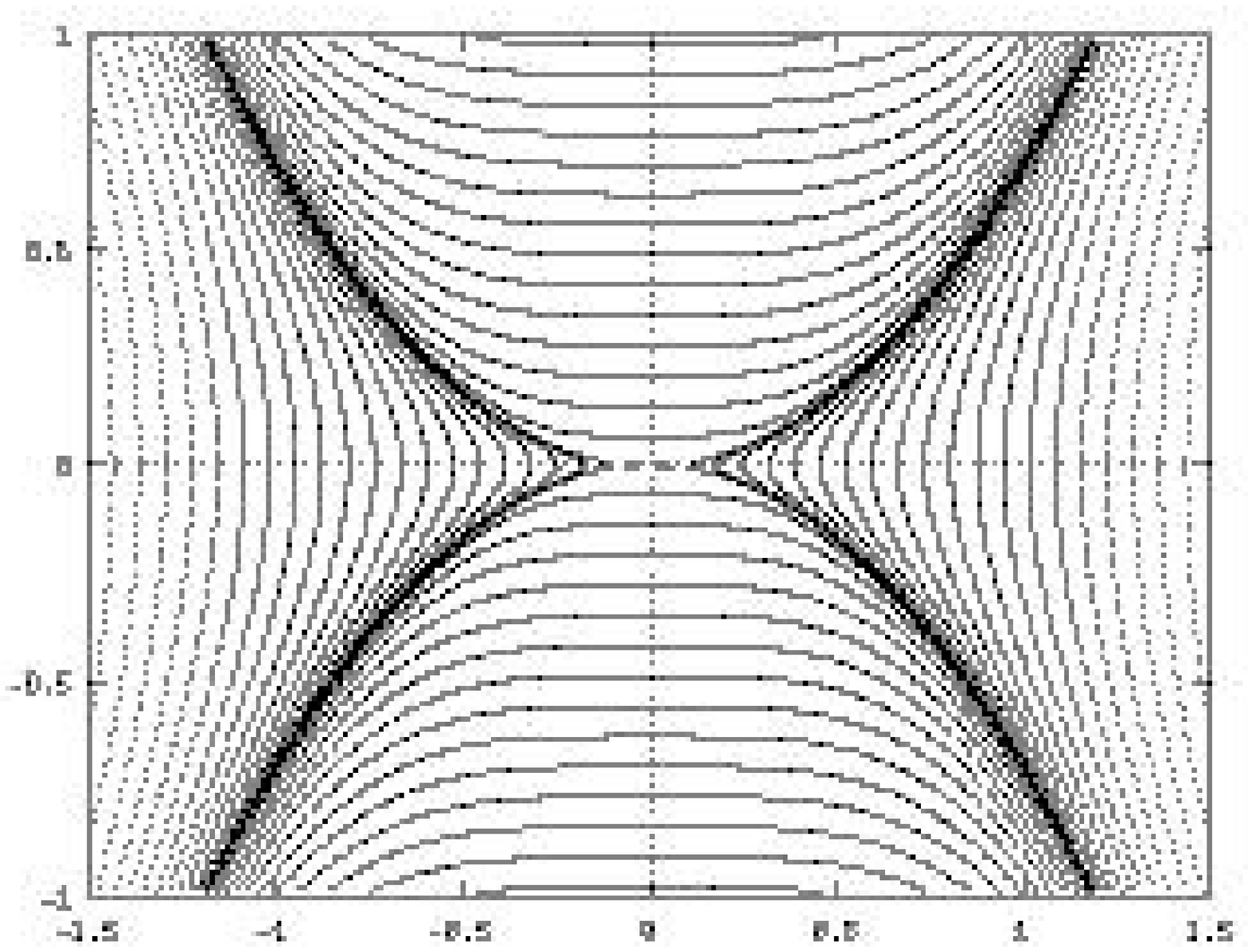}
\includegraphics[width=5cm]{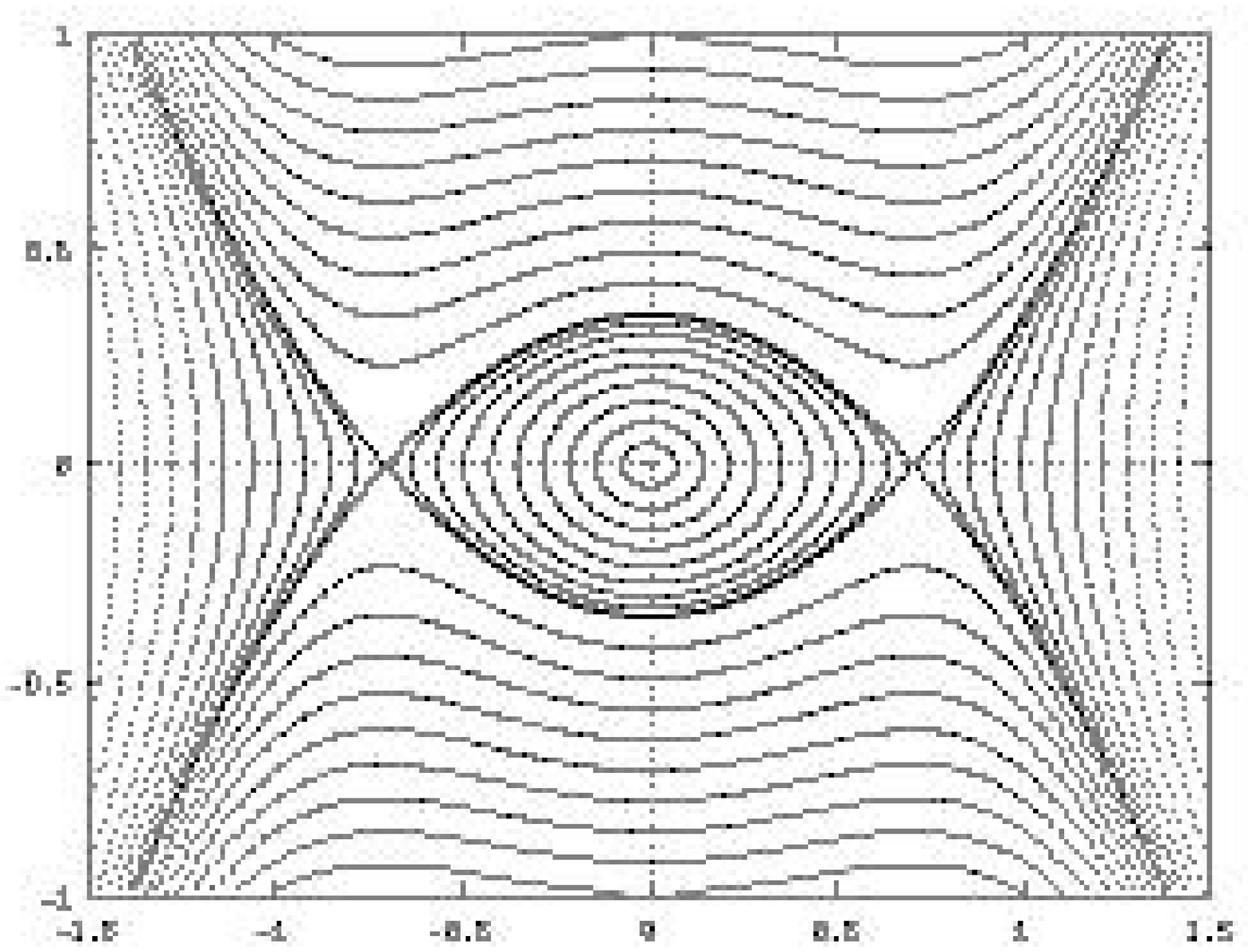}
}
\caption{Lines of the constant energy on the phase space $(u,v)$ 
for the period-doubling bifurcation $(D = -1)$, for $\eps = -0.1$, $\eps = 0$, $\eps  = 0.1$}
\label{fig:PDphaseN}
\end{figure}

The normal form of a Hamiltonian system with two degrees of freedom near a 
periodic orbit in $1:2$ resonance is
\begin{equation} \label{Hpd}
    H(u,v)=\frac{v^2}{2}+D\frac{u^4}{4}+\eps u^2\,,
\end{equation}
where $D = \pm 1$.
As in the case of the saddle-centre bifurcation the variables $u$ and $v$ are canonically conjugate variables
on a local transversal Poincar\'e section and $\eps$ is a parameter, typically corresponding to
the energy of the original system. Contourplots of this Hamiltonian
show the intersection of invariant tori with the Poincar\'e section transversal 
to the bifurcating orbit.
For $D=1$ they are shown in Fig.~\ref{fig:PDphaseP}, 
for $D=-1$ in Fig.~\ref{fig:PDphaseN}, 
for $\eps < 0$, $\eps = 0$, and $\eps > 0$, respectively.

\begin{figure}
\centerline{\includegraphics[width=12cm]{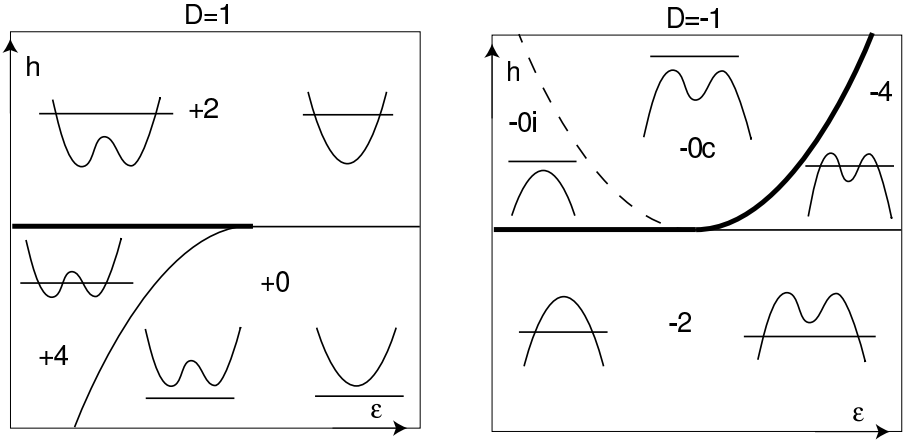}}
\caption{Schematic sketch of the bifurcation diagram for the two cases of 
the period doubling bifurcation.
Graphs of $P_4$ are shown together with a horizontal line indicating the value of $h$.
The bold line denote critical values of unstable orbits. 
The dashed line is not critical; it indicates a vanishing of the discrimiant.
} \label{fig:PDbif}
\end{figure}

The critical points and critical values of the energy map $H : \R^2 \to \R$ and their
dependence on the parameter $\eps$ describe the structure of the bifurcation.
The Hamiltonian has critical points at $(0,0)$ and at $(u,v) = (\pm\sqrt{-2\eps D},0)$
with critical values $h=0$ and $h=-D\eps^2$. The origin is a local minimum 
for $\eps>0$, a saddle otherwise. The second critical point 
exists when $\eps D < 0$, and is a local minimum for $\eps<0$, a saddle otherwise.
The set of critical values, see Fig.~\ref{fig:PDbif}, therefore is the union of the line $h=0$ with 
the half of the parabola $h = -D\eps^2$ for which $D\eps < 0$.
For $D>0$ the only unstable branch in the bifurcation diagram is
$h=0$ for $\eps < 0$. It divides the two regions of real motion.
A third region $\{h < -\eps^2\} \cup \{\eps > 0, -\eps^2 < h < 0\}$ is not 
in the image of $H$, hence there is no real motion corresponding
to $(h,\eps)$ from this region.

For $D<0$ the line of critical values $h=0$ again has a saddle 
as critical point when $\eps < 0$. In addition the half parabola
$h=\eps^2$, $\eps > 0$ also corresponds to a saddle of $H$ that
is not at the origin. When $D< 0$ the whole plane $(h,\eps)$ is 
in the image of $H$ considering all $\eps$. The critical values divide
the plane into three regions with 0, 2, and 4 real roots.

The dynamics of the reduced one degree of freedom system is given by
$\dot u = v$ and eliminating $v$ using the Hamiltonian gives a first
order equation for $u$. Separation of variables then gives the period
of motion in the reduced one degree of freedom system as the elliptic integral
\[
    T(h,\eps)=\oint\frac{{\rm d}u}{\sqrt{2h - \frac12 D u^4-2\eps u^2}}\,.
\]
The number of parameters could be reduced by introducing the ratio $h/\eps^2$,
but for clarity we do not introduce this scaling.
The period $T$ is an elliptic integral on the curve
\begin{equation} \label{ElliCurve4}
  {\cal E}: w^2 = P_4(z) = P_2(z^2) =  2h - \frac12 D z^4 - 2\eps z^2  \,.
\end{equation}
The discriminant of $P_4$ is
\[
\Delta = -256 D h (D h + \eps^2)^2\,.
\]
It vanishes for $h = 0$ and $Dh  = -\eps^2$, corresponding to the critical 
values $h=0$ and $h= -D \eps^2$ already found above. 
As usual the set of critical values is contained in the discriminant, however,
the discriminant vanishes on a larger set. Namely the branch of the parabola
with $D\eps > 0$ is not part of the critical values.
The three regions already found correspond to regions with 
 4, 2, and 0 real roots of $P_4$, respectively, see Fig.~\ref{fig:PDbif}. 
 The part of the discriminant that is not part of the critical values is dashed.
The polynomial can be factored as
 \[
     P_4(z) = -\frac12 D(\xi_- - z^2 )(\xi_+ - z^2 ), \quad
     \frac12 D\xi_{\pm} =  - \eps \pm \sqrt{\eps^2 + Dh}\,.
 \]
 Comparing coefficients gives $\xi_-\xi_+ = -4hD$ and $\xi_- + \xi_+ = -4\eps D$. 
The factorization of $P_4$ has real factors, i.e.\ $\xi_\pm \in \R$,  with one exception. 
It occurs when the quadratic equation $P_2(\xi)= 0$ has complex roots.
The position of the roots in the regions of the parameter plane is
 as follows.
 The two real cases for $D = 1$ are
\begin{align*}
+2:  && h > 0                                   &&  \xi_- < 0 < \xi_+ && 2 [-\sqrt{\xi_+},\sqrt{\xi_+}] \\ 
+4: && \eps < 0, -\eps^2 < h < 0  &&  0 < \xi_- < \xi_+ && 2 [\sqrt{\xi_-},\sqrt{\xi_+}] 
\end{align*}
In the first column a label is given containing the number of real roots.
All 4 cases appear when $D=-1$ for
\begin{align*}
-2:    && h < 0                                  &&  \xi_+ < 0 < \xi_-  && 2[-\infty, -\sqrt{\xi_-} ] \\
-4:   && \eps > 0, 0 < h  < \eps^2 &&   0 < \xi_+ < \xi_- && 2[-\infty, -\sqrt{\xi_-} ] \\
-0i: && \eps < 0, 0 < h <  \eps^2 &&   \xi_+ < \xi_- < 0 && [-\infty,\infty] \\
-0c: && h > \eps^2                         && \xi_\pm \text{ complex} && [-\infty,\infty] 
\end{align*}
The last column gives the interval of real motion along which the period integral is taken.
The region without real roots contains two parts separated by the branch 
of the parabola $h = -D\eps^2$ with $D\eps > 0$. 
On this branch there occurs a collision of complex roots at $z^2 = -2\eps D$,
and they move from the imaginary axis into the complex plane.

The intervals of integration and multiplication factors as given in the last
column of the previous table can be read off from
the phase diagrams, see Fig.~\ref{fig:PDphaseP}, \ref{fig:PDphaseN}. 
In the case $D=-1$ with 4 real roots 
there are different orbits for the same $(h,\eps)$. One is the compact orbit 
near the stable fixed point with extent $2[-\sqrt{\xi_+}, \sqrt{\xi_+} ]$. The interval
given above is for the non-compact orbit, which has half of this period.

In order to calculate the derivative of $T$ with respect to $h$ the integral
is first written in standard form and then differentiated. This more traditional
approach (as compared to the previous section) is preferable in this case 
because the roots of the quartic are easily written down.

\begin{figure}
\centerline{\includegraphics[width=9cm]{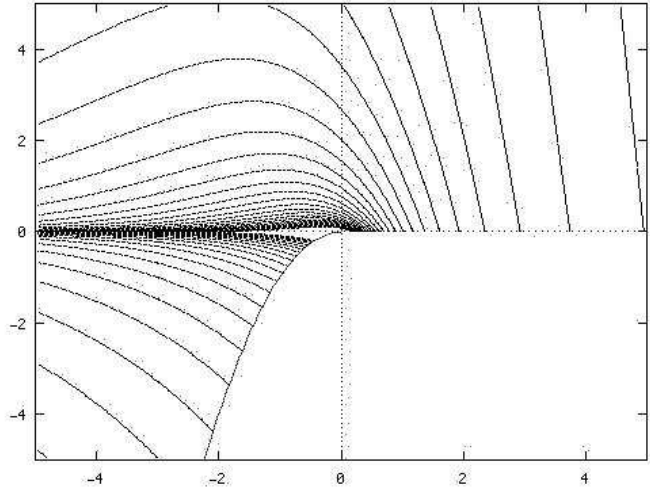}}
\caption{Lines of constant period $T=1/\Omega$ equidistant with $\Delta T = 0.3$ 
on the parameter plane $(\eps, h)$ for the period-doubling bifurcation ($1:2$ resonance, 
$\omega = 1/2$) with $D = 1$}
\label{fig:PDWconstP}
\end{figure}

\begin{figure}
\centerline{\includegraphics[width=9cm]{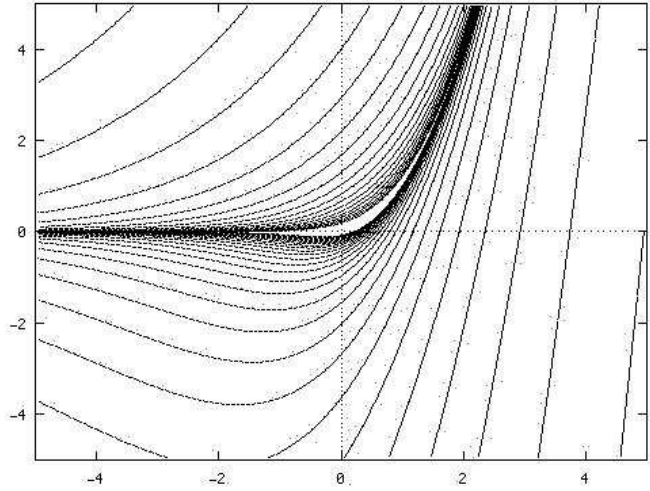}}
\caption{Lines of constant period $T=1/\Omega$ equidistant with $\Delta T = 0.3$ on the 
parameter plane $(\eps, h)$ for the period-doubling bifurcation 
($1:2$ resonance, $\omega = 1/2$) with $D = -1$}
\label{fig:PDWconstN}
\end{figure}

Denote the ratio of the roots $\xi_\pm$ by 
\[
   r = \frac{\xi_-}{\xi_+} = \frac{-\eps - \sqrt{\eps^2 + Dh}}{-\eps + \sqrt{\eps^2 + Dh}} \,.
\]
Then the period $T(h,\eps)$ in the 6 cases is given by 
\begin{align}
 +2,-2: && T = & \left(\frac{8(1-2k^{2})}{\eps}\right)^{1/2} K(k), 
          && k^2 =  \frac{1}{1 - r}, \\
 +4,-0i: && T = &  \left(\frac{8(k^{2}-2)}{\eps}\right)^{1/2} K(k),
          && k^2 =  1 - r, \\
 -4: && T = & \left(\frac{8(1+k^{2})}{\eps}\right)^{1/2} K(k), 
          && k^2 =  \frac{1}{r}, \\
 -0c: && T = &  \left(\frac{8(2k^{2}-1)}{\eps}\right)^{1/2} K(k), 
           && k^2 = \frac12 + \frac{\eps}{2\sqrt{h}}\,.
\end{align}
The level lines of the period $T$ (and hence the rotation number $\Omega = 1/T$) are shown in 
Fig.~\ref{fig:PDWconstP} for $D=1$ and in 
Fig.~\ref{fig:PDWconstN} for $D=-1$.
These numerically computed pictures show that there are no 
vertical tangents, hence the twist does not vanish. 
This is now proved by differentiating the period in Legendre 
normal form.

The twist vanishes when 
\[
  0 = \frac{\partial T}{\partial h}  = \frac{\partial T}{\partial k}  \frac{\partial k}{\partial h} \,.
\]
The last factor $\partial k/\partial h$ does not vanish for  $k\in (0,1)$.
In the cases $+2$, $-2$, and $-0c$ it seems to vanish when $k=1/2$. 
But this implies that $\eps = 0$ and the singularity cancels.
Therefore it is enough to consider ${\partial T}/{\partial k} = 0$.
After removing common non-vanishing factors the conditions for 
vanishing twist are $V_i(k) = 0$, $i=1,2,3$ where
\begin{align}
+2, -2, -0c: && V_1(k) =& (1-2k^2) E(k)   - (1-k^2) K(k),  \\
+4, -0i: && V_2(k) =& \left(1-\frac{k^2}{2}\right) E(k) - (1-k^2) K(k), \\
-4:  && V_3(k) =& (1+k^2) E(k)   - (1-k^2) K(k) \,.
\end{align}
The three equalities $V_i(k) = 0$, $i=1,2,3$ are never satisfied on the range $k \in  (0,1)$.
Obviously $V_i(0) = 0$, while $V_i(1) = -1, 1/2, 2$ for $i=1,2,3$, respectively. So 
we need to show that $V_1$ is negative and $V_2$, $V_3$ are positive for 
$k \in (0,1)$.
Differentiating $V_1$ and $V_2$ gives the simple results
\begin{align}
V'_2(k)  = & -\frac32k (K(k) - E(k)) \\
V'_3(k)  = & -3k E(k)
\end{align}
In the last case $V'_3(k)$ is obviously non-positive, so that the twist in the case
$-4$ is a monotone funtion rising from 0, hence it is nonzero.
Similarly  also $V'_2(k)$ is non-positive, which follows from 
the well known inequality $K(k) > E(k)$. 
The first function $V_1(k)$ is negative on $(0,1)$, but not monotone.
Rewriting it as 
\[
  V_1(k) = (1-k^2)(E(k)-K(k)) - k^2 E(k) < 0
\]
the inequality is clear because both terms are negative for $k\in (0,1)$.
Therefore the twist never vanishes in a neighborhood of 
the period doubling bifurcation.

The relation between the cases $D= \pm1$ is interesting. The main observation is
that changing the sign of $D$ and $\eps$ inverts the overall sign of the potential
$V(x;D,\eps) = Dx^4/4 + 2\eps x^2$. Therefore changing the signs of $D$, $\eps$,
and $h$ leaves the roots of $P_4$ invariant. 
By this mapping the regions in parameter space with the same numbers of real
roots are mapped into each other. The integration path does change in a 
less trivial way. The integration needs to be taken over the positive intervals
of $P_4$ on the real axis. Changing the sings of $D$, $\eps$, and $h$ does change
the sign of $P_4$.
As a result the periods for 
$D=1$ are the $\alpha$ cycles of the elliptic curve, while those for $D=-1$ are
the $\beta$ cycles. Hence the period for the case $-4$ can be obtained
from that of $+4$ by replacing $K(k)$ by $K'(k) = K(\sqrt{1-k^2})$.
Similarly, the cases $+2$ and $-2$ are mapped into each other.

The obvious symmetry in Figures~\ref{fig:PDWconstP} and \ref{fig:PDWconstN}
with respect to changing the sign of $h$
is related to the fact that changing the sign of $D$ and $h$ leaves the 
ratio $r$, and therefore also the corresponding modulus $k^2$, invariant.
This means that the level lines in region $+2$ and $-2$ can be obtained from 
each other by reflection on the $h=0$ axis. In a similar way the regions $+4$ and 
$-0i$ have the same rotation number. 

The period doubling could have been treated in a scaled version with only 
one essential parameter $\delta = h/\eps^2$. However, the presentation seems
more transparent in the unscaled version. Similar to the case of the saddle-centre
bifurcation the modulus $k^2$ of the elliptic integral for the period is constant
on the parabolas $h = \delta \eps^2$. Again dependence on the parameters on 
these curves is simply algebraic, as before through $|\eps|^{1/4}$. 
The value of the modulus is not defined at the origin, but depends on the 
parabola on which it is approached. But in any case, there are no twistless tori
near the origin in this bifurcation.

\section{Universality} \label{sec:Universality}

A major problem in our approach seems to be that the invariant twistless tori 
in the normal form are not compact. The normal form is obtained from
an expansion near the bifurcation point, and is therefore local in phase space
and local in the parameter. How can the rotation number of a non-local 
invariant torus be determined from this local normal form? 
To answer this question higher order terms need to be considered in 
the Hamiltonian. The Poincar\'e map near the bifurcation can be described by 
the time $1$ map of the Hamiltonian 
\[
 \tilde H(u,v,t;\eps) = H(u,v;\eps) + G(u;\eps) + R(u,v,t;\eps)\,.
\]
The remainder terms $R$ containing the periodic time dependence can 
be made arbitrarily small, but they cannot in general be removed while 
retaining a non-zero radius of convergence of the normal form. 
The first term $H$ is the normal form analysed in the previous chapters. 
We will concentrate on the saddle-centre case (\ref{Hsn}), 
but similar remarks apply to (\ref{Hpd}).
The invariant tori near the origin for $\eps>0$ of $H$ are not compact. 
The higher order terms in $G$ can compactify them. The results about
vanishing twist can be applied when $G$ does compactify these curves.
However, the precise form of $G$ does not matter. 
Under the compactness assumption KAM theory can be applied to $H+G$, 
where $R$ is the perturbation. Many of the invariant tori of $H+G$ will persist. 
In particular a twistless invariant torus of $H+G$ will persist if it is 
sufficiently irrational. The curve (\ref{VTg0curve}) in the parameter plane
therefore does not have invariant twistless tori in its perimage for
every point, instead just for a cantorset of points. This is well understood. 
The main issue in this section is to understand the effect of adding $G$ to $H$.
For small $\eps$ the essential contribution to the
diverging period comes from the dynamics near the origin, while
the dynamics on the invariant torus away from the origin has finite period.
This is why the local normal form can give a statement
about the dynamics on a non-local invariant torus near the bifurcation point.
In particular we will now show that the curve of vanishing twist that was found 
to be emanating from the cusp of the saddle-centre bifurcation has 
a universal shape sufficiently close to the cusp singularity. 
In particular this means that the constant $\gamma_0$ that determines the
shape of the curve of twistless tori in relation to the curve of critical values of the
unstable orbits has the universal value $\gamma_0 \approx  0.91522$.
Quantities derived from $\gamma_0$, like $z_0$ and the coefficients in (\ref{Omlim}),
are accordingly also universal.
Our calculation will show that the value of $\gamma_0$ is not influenced by the higher order
terms $G$ in the Hamiltonian. Moreover, the following argument also shows that
 integrating the non-compact invariant torus up to $v = \pm \infty$ does not introduce 
 and additional error.

Let the high-order truncated normal form Hamiltonian be
\[
 H(u,v;\eps) = \frac12 v^2 + \frac13 u^3 + \eps u + G(u,\eps)
\]
where $G(0,\eps) = 0$ and $G(u,0) = 0$ is an analytic function. 
It is not necessary to assume that the higher order terms in $G$ depend
on $v$ also, see e.g.~\cite{MH92}, but even with such a dependence a slightly 
modified argument would work. As already explained, it is now assumed that 
$G(u,\eps)$ is such that the invariant curves near the origin for 
$\eps > 0$ are compact.

Hamiltons equation for $u$ reads $\dot u = v$ as before.
The period is obtained by solving $H(u,v;\eps) = h$ for $v = v(u;h,\eps)$
and then by integrating 
\[
     T(h,\eps) = \oint \frac{{\rm d} u}{v(u;h,\eps)} \,.
\]
The idea is to split the integral into two parts; one part near the origin, where the 
main contribution originates, and the rest, which is called $T_3$. 
In addition the singular integral near
the origin is split again into two parts, $T_1$ which has the same integrand 
as in the previous calculation with $G=0$, and a correction $T_2$ which contains
the contribution from $G$. The integral $T_1$ will be the most singular, 
$T_2$ is mildly singular, and $T_3$ is regular. For sufficiently small $\eps$ 
then $T_1$ dominates and the previous result is recovered. 

To achieve the splitting into $T_1$ and $T_2$ the multiplicative structure of 
the original inegrand has to be recovered. Solving $H=h$ for $v$ and
inserting into $\dot u$ gives
\[
    \dot u^2 = v(u;h,\eps)^2 = Q_0(u;h,\eps) \hat Q(u;h,\eps)
\]
where $\hat Q = 1 + O(u)$ when $|u| \le C \le 1$ for some fixed constant $C$,
and $Q_0$ is a polynomial of degree 3 in $u$ whose zeroes approach those
of the original case with $G=0$ when $\eps \to 0$. Denote by $\gamma_C$ the part of 
$H(u,v;\eps) = h$ for which $|u| < C$, and $\bar \gamma_C$ the rest of the 
invariant torus. Then the period integral can be split as
\begin{align}
T(h,\eps) = & \int_{\gamma_C} \frac{\dee u}{\sqrt{Q(u,\eps,h)}} + 
                        \int_{\bar \gamma_C} \frac{\dee u}{\sqrt{Q(u,\eps,h)}} \\
                   = & \int_{\gamma_C}  \frac{\dee u}{\sqrt{Q_0(u,\eps,h)}} + 
                          \int_{\gamma_C}  \frac{{\hat Q(u,h,\eps)}^{-1/2} - 1}{\sqrt{Q_0(u,\eps,h)}} \, \dee u + 
                          \int_{\bar \gamma_C} \frac{\dee u}{\sqrt{Q(u,\eps,h)} }        \\
                   = & T_1 + T_2 + T_3 \,.                 
\end{align}
Now $T_3$ is regular, and gives a bounded contribution, so we can ignore it.
The integral $T_1$ is singular in the limit $h,\eps \to 0$. $T_2$ is less 
singular, and in particular bounded, because the numerator goes to zero in this limit. 
So we only need to show that $T_1$ approaches the complete integral when 
$\eps \to 0$. The integral is
\[
   T_1(h,\eps) = \frac{4\sqrt{3/2}}{r^{1/2}} F(\phi,k), \quad
   \phi = 2 \arctan \sqrt{\frac{u_0+C}{r}} \,, 
\]
where $u_0$ is the real root, $\zeta$ is the complex roots of $Q_0$, 
and $r$ and $k$ are as before,
\[
   r = \sqrt{2u_0^2 + |\zeta|^2}, \quad
   2k^2 = 1 + \frac{3u_0}{2r}  \,.
\]

The integral $T_1$ does not diverge because of the modulus $k^2 \to 1$,
but because in the prefactor $r \to 0$.
In fact, we already observed that the modulus $k^2 \le (\sqrt{3} + 2)/4 < 1$
inside the first quadrant $\eps \ge 0, h > 0$. 
The behaviour of $r$ for small $\eps$ is obtained from 
\[
   r^2 = \frac32 u_0 \sqrt{1 + \alpha^2} = D(\gamma) \eps^{1/4} \,.
\]
This shows that $\phi \to \pi$, and the integral approaches the 
complete integral with the same modulus $k(\gamma)$ as before.
Therefore even though $T_1$ is an incomplete integral, in the limit
of small $\eps$ its value approaches that of the complete integral 
$T(h,\eps)$, which diverges. Since the other integrals $T_2$ and $T_3$
are finite the analysis obtained from the complete integral over the
non-compact curve of $H(u,v;\eps)$ with $G=0$ therefore correctly
describes the behaviour of the rotation number of the compact invariant
curve obtained when $G \not = 0$.

\goodbreak

\section{Example: H\'enon Map}

\begin{table}
\centerline{
\begin{tabular}{lll}
$\eps$ & $u \eps^{-1/2}$ & $\Omega \eps^{-1/4}$ \\ \hline
0.1    & 0.027 &  0.28314 \\
0.01    & 0.328 &  0.17278 \\
0.001   &  0.478 &  0.14656 \\
0.0001   & 0.528 &  0.14008 \\
0.00001  &  0.545 & 0.13823 \\
0.000001 &  0.550  & 0.13767
\end{tabular}
}
\caption{Numerically measured position $u$ and rotation number $\Omega$ of the
twistless curve for 
the three times iterated H\'enon map with parameter $k = 1 - \eps$.} \label{tabHen}
\end{table}

The H\'enon map in the area preserving case,
\[
      (x', y') = (y - k + x^2, -x) \,,
\]
illustrates the above. 
There are saddle-centre bifurcations in the H\'enon map  for which the invariant tori
for $\eps > 0$ are not compact, and hence the vanishing twist cannot be observed.
This applies to the initial bifurcation at $k=-1$ that creates the pair of 
fixed points, and also to many saddle-centre bifurcations that occur for $k > 4$. 
However for $k=1$ a pair of period 3 orbits is created at 
the origin in a saddle-centre bifurcation of the third iterate of the map, 
and the corresponding invariant tori are compact.
One of the period three points is located on the  symmetry line $x = -y$. 
In new coordinates $(u,v) = (x, y+x)$ the third  iterate of the map expanded 
near the origin with parameter $k = 1- \eps$ is
\[
  (u', v') =  \left( (u-v)(1 + 4v) , v + \eps + (u^2 - 2uv + 3v^2)\right) + O(u^3,v^3,\sqrt{|\eps|}^3)\,.
\]
The (exact) location of the fixed points is $(u,v) = ( \pm \sqrt{-\eps}, 0)$, 
with trace of the Jacobian $2 \mp 2 \sqrt{-\eps} + 8\eps \mp 8 (-\eps)^{3/2}$
so that the (approximate) multiplier is $\mu \approx 1 + \imag \sqrt{2} (-\eps)^{1/4} $
and $\bar\mu$ for the fixed point at $u = \sqrt{-\eps}$. The corresponding 
rotation number $\omega$ of the elliptic fixed point 
is obtained from $\mu = \exp( 2\pi \imag\omega)$, so 
that $\omega = (-\eps)^{1/4} / (\sqrt{2}\pi)$.
For small positive $\eps$ the H\'enon map possesses compact invariant curves
near the origin. A higher order normal form would give a $G$ that describes 
these invariant curves. The H\'enon map is non-integrable, so that only 
sufficiently irrational invariant curves of the (high order) normal form will
exist in the H\'enon map. For the situation under consideration numerical 
experiments show that many of these invariant curves do exist.

Considering the third iterate of the H\'enon map turns the pair of period three orbits
into three pairs of fixed points with heteroclinic connections. In the normal form there
is only one pair of fixed points, and the unstable fixed point has a homoclinic 
connection. To match the prediction in the case of more than one unstable fixed
point the period and hence rotation number must be calculated for the heteroclinic
connection. For $\eps > 0$ there are invariant tori on which the dynamics becomes 
slow near the three points that are close to the three bifurcation points of the third 
iterate of the map. 
Hence the rotation number for the third iterate
of the map between two successive such points gives the correct rotation number.
The results together with the position of the twistless curve are shown in 
table~\ref{tabHen}. The values shown converge to the predicted values
given in (\ref{z00}) and (\ref{Omlim}),
however, fairly small $\eps$ are needed to see this.
The convergence to the true value $0.1374244...$ occurs 
approximately as $\tfrac12 \eps^{5/4}$.

\bibliographystyle{plain}
\bibliography{notwist}

\def\cprime{$'$}
\begin{thebibliography}{10}

\bibitem{Arnold78}
V.~I. Arnold.
\newblock {\em Mathematical Methods of Classical Mechanics}, volume~60 of {\em
  Graduate Texts in Mathematics}.
\newblock Springer, Berlin, 1978.

\bibitem{CGM96}
D.~del Castillo-Negrette, J.M. Greene, and P.J. Morrison.
\newblock Area preserving nontwist maps: Periodic orbits and transition to
  chaos.
\newblock {\em Physica D}, 91(1):1--23, 1996.

\bibitem{DL98}
A.~Delshams and R.~de~la Llave.
\newblock K{AM} theory and a partial justification of {G}reene's criterion for
  nontwist maps.
\newblock {\em SIAM J. Math. Anal.}, 31(6):1235--1269 (electronic), 2000.

\bibitem{DM02}
H.~R. Dullin and J.~D. Meiss.
\newblock Twist singularities for symplectic maps.
\newblock {\em Chaos}, 13(1):1--16, 2003.

\bibitem{DMS98b}
H.~R. Dullin, J.~D. Meiss, and D.~G. Sterling.
\newblock Generic twistless bifurcations.
\newblock {\em Nonlinearity}, 13:203--224, 2000.

\bibitem{HowHum95}
J.~E. Howard and J.~Humpherys.
\newblock Nonmonotonic twist maps.
\newblock {\em Physica D}, 80(3):256--276, 1995.

\bibitem{HowHoh84}
J.E. Howard and S.M. Hohs.
\newblock Stochasticity and reconnection in hamiltonian systems.
\newblock {\em Physical Review A}, 29:418, 1984.

\bibitem{MH92}
K.~R. Meyer and G.~R. Hall.
\newblock {\em Introduction to Hamiltonian Dynamical Systems and the N-Body
  Problem}.
\newblock Springer, Berlin, 1992.

\bibitem{Moeckel90}
R.~Moeckel.
\newblock Generic bifurcations of the twist coefficient.
\newblock {\em Ergodic Theory Dynam. Systems}, 10(1):185--195, 1990.

\bibitem{Sadovskii96}
D.~A. Sadovskii and J.~B. Delos.
\newblock Bifurcation of the periodic orbits of hamiltonian systems: An
  analysis using normal form theory.
\newblock {\em Phys. Rev. E}, 54:2033--2070, 1996.

\bibitem{SM71}
C.L. Siegel and J.K. Moser.
\newblock {\em Lectures on Celestial Mechanics}.
\newblock Classics in Mathematics. Springer-Verlag, New York, 1971.

\bibitem{Simo98}
C.~Sim{\'o}.
\newblock Invariant curves of analytic perturbed nontwist area preserving maps.
\newblock {\em Regular {\&} Chaotic Dynamics}, 3:180--195, 1998.

\bibitem{WVCP88}
J.~P. Van~der Weele and T.~P. Valkering.
\newblock The birth process of periodic orbits in nontwist maps.
\newblock {\em Physica A}, 169(1):42--72, 1990.

\end{thebibliography}

\end{document}